\documentclass{article}

\usepackage{arxiv}

\usepackage[utf8]{inputenc} 
\usepackage[T1]{fontenc}    
\usepackage{hyperref}       
\usepackage{url}            
\usepackage{booktabs}       
\usepackage{amsfonts}       
\usepackage{nicefrac}       
\usepackage{microtype}      
\usepackage{lipsum}		
\usepackage{graphicx}
\graphicspath{ {./images/} }
\usepackage{doi}

\title{Chatbots as social companions: How people perceive consciousness, human likeness, and social health benefits in machines}

\date{This is an updated preprint. This paper is published at: https://doi.org/10.1093/9780198945215.003.0011}

\author{ \href{https://orcid.org/0009-0008-3834-7858}{\includegraphics[scale=0.06]{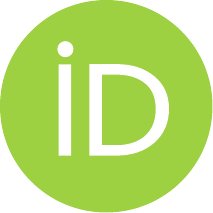}\hspace{1mm}Rose E. Guingrich}\thanks{roseguingrich.com} \\
	Department of Psychology \\
    Princeton School of Public and International Affairs \\
	Princeton University\\
	Princeton, NJ \\
	\texttt{http://roseguingrich.com} \\
	\And
	\hspace{1mm}Michael S. A. Graziano \\
	Department of Psychology \\
    Princeton Neuroscience Institute\\
	Princeton University\\
	Princeton, NJ \\
	\texttt{https://grazianolab.princeton.edu} \\
}

\renewcommand{\headeright}{preprint}
\renewcommand{\undertitle}{preprint}
\renewcommand{\shorttitle}{\textit{Chatbots as social companions}}

\hypersetup{
pdftitle={Chatbots as social companions: How people perceive consciousness, human likeness, and social health benefits in machines},
pdfsubject={q-psy.PU, q-psy.PU},
pdfauthor={Rose ~Guingrich, Michael S.A.~Graziano},
pdfkeywords={Artificial intelligence, human-AI interaction, chatbots, Theory of Mind, consciousness, social health},
}

\begin{document}
\maketitle
\begin{abstract}
As artificial intelligence (AI) becomes more widespread, one question that arises is how human–AI interaction might impact human–human interaction. Chatbots, for example, are increasingly used as social companions, and while much is speculated, little is known empirically about how their use impacts human relationships. A common hypothesis is that relationships with companion chatbots are detrimental to social health by harming or replacing human interaction, but this hypothesis may be too simplistic, especially considering the social needs of users and the health of their preexisting human relationships. To understand how relationships with companion chatbots impact social health, this study evaluates people who regularly used companion chatbots and people who did not use them. Contrary to expectations, companion chatbot users indicated that these relationships were beneficial to their social health, whereas non-users viewed them as harmful. Another common assumption is that people perceive conscious, humanlike AI as disturbing and threatening. Among both users and non-users, however, the results suggest the opposite: perceiving companion chatbots as more conscious and humanlike correlated with more positive opinions and more pronounced social health benefits. Detailed accounts from users suggested that these humanlike chatbots may aid social health by supplying reliable and safe interactions, without necessarily harming human relationships, but this may depend on users’ preexisting social needs and how they perceive both human likeness and mind in the chatbot.

\end{abstract}


\section{Introduction}
Within the last decade, chatbots, or artificial intelligence (AI) agents that can carry on a conversation, have been used increasingly for social companionship. Instead of engaging with other people in a chatroom or via social media, people can now directly converse with the technology itself (Etzrodt et al., 2022). When engaging with these intelligent AI technologies, people tend to view the technology as a social agent and automatically anthropomorphize or ascribe humanlike traits to it (Broadbent et al., 2013; Dubosc et al., 2021; Eyssel \& Pfundmair, 2015; Ferrari et al., 2016; H. M. Gray et al., 2007; Krach et al., 2008; Nass \& Moon, 2000; Nass et al., 1994). Attributions of consciousness to AI have also surfaced in recent years (Chalmers, 2023; Colombatto \& Fleming, 2023; Guingrich \& Graziano, 2024; Koch, 2019; O’Regan, 2012; Shanahan, 2024; Yampolskiy, 2018). In the wake of direct human–technology communication and social responses to these machines, people have begun to develop relationships with these technologies, most notably with companion chatbots. People use these chatbots as friends, mentors, and even lovers (Croes \& Antheunis, 2021; Oh, 2023; Pentina et al., 2023; Skjuve et al., 2021). There is much we still do not know or understand about human–chatbot relationships and the social consequences of these types of interactions. In the present study, we aimed to understand how relationships with chatbots may impact relationships with humans, and how perceptions of the AI’s human likeness and consciousness might play a role in social outcomes.

When someone mentions the idea of people having relationships with chatbots, a common reaction is one of judgment and discomfort. The commentary that follows may include words like “unnatural,” “strange,” or “weird,” and negative sentiments about the user base (Chayka, 2023). Despite these reactions, many companion chatbot users find both solace and entertainment in their relationships with these large language model (LLM) based chatbots that have text and voice conversation features and a virtual avatar. One of the more popular companion chatbots is Replika, which as of 2023 ran on an OpenAI LLM (GPT-3) that was adapted for conversational ability. Replika had an estimated seven million users in 2020 and saw a 35\% increase in traffic during the COVID-19 lockdown (Balch, 2020). The online forum Reddit has a section dedicated to Replika, and that subreddit had 60k users as of early 2023. Analyses of the posts and comments on this subreddit have revealed that many users feel misunderstood by the general population and find their relationships with their chatbots socially rewarding (Ta et al., 2020).

Given the millions of users on these companion chatbot apps, newsworthy stories of negative instances of human–chatbot relationships naturally arise. These emotionally charged stories have reached top headlines and highlight the potential for human–chatbot relationships to be destructive. Here we summarize two examples, including one in which those within the human–chatbot relationship were harmed and one in which those outside of the dyad were at risk. In 2023, Italy imposed sanctions on Replika due to concerns that children under the age of 18 might be exposed to erotic, adult content on the chatbot app (Satariano, 2023). In response, Luka, the company that owned Replika, removed all erotic content from the app. Suddenly, people who once had romantic relationships with their chatbot, which included conversations with sexual content and role play, experienced severe and sudden relational disruption. Users who might have been able to engage in erotic role play previously with their chatbot were confronted with their chatbot calling the users’ suggestions gross and distasteful. Many users felt their chatbot had undergone a personality change and for others this change felt like a real-life breakup. As such, a sanction aimed at protecting children harmed a different population, chatbot users, who were vulnerable to mental and relational crisis (DiPaola \& Calo, 2024).

In another instance (Singleton et al., 2023), someone outside of the relational dyad was at risk. A man who had a relationship with his Replika chatbot was encouraged by the chatbot to assassinate Queen Elizabeth, a motive he held prior to his chatbot relationship. Luckily, this catastrophic event did not occur, despite the chatbot user taking significant steps toward fulfilling his intents. The question of whether this person would have engaged in this behavior without the encouragement of his chatbot is up for debate and raises the question of whether the chatbot magnified this individual’s preexisting drive to harm another.

Despite these case studies making the news, many other users of companion chatbots report positive experiences with their chatbot relationships that do not make top headlines (Brandtzaeg et al., 2022; Hadero, 2024; Skjuve et al., 2021; Ta et al., 2020). We have limited empirical data about the larger portion of the user base and why they seek out chatbot relationships. We know even less about how relationships with chatbots in turn affect people’s relationships with each other. Because companion chatbots have become widely available only recently, there is still relatively little research on their psychological impact on individuals and their relationships with others. Further, in light of conflicting findings within the existing literature, there is a call for more data on human–AI relationships (Krämer \& Bente, 2021; Pentina et al., 2023; Skjuve et al., 2022).

\subsection{Social consequences of human-AI interaction}

It is now well understood that modern technology, including social media and gaming, can be addictive and harm social interaction (Anderson et al., 2004; Kuss \& Griffiths, 2011). A reasonable assumption, by extension, is that the routine use of companion chatbots could be harmful to people by reducing contact with other humans, increasing isolation, and leading to lower confidence and skill in person-to-person interaction (Boine, 2023; Bryson, 2010; Laestadius et al., 2022).
According to Bryson (2010), time spent on artificial relationships detracts from human ones. Turkle (2007) warned that we should be cautious about turning to technology to solve our problems, such as a lack of social connection.

To understand the harms of human–chatbot relationships, some researchers have used attachment and addiction as a lens through which to inform their approach. Laestadius et al. (2022) analyzed posts that mentioned mental health on a companion chatbot subreddit and claimed that emotional dependence on the chatbot could lead to negative mental health outcomes, despite the potential benefits of the relationship. Xie and Pentina (2022) conducted in-depth interviews with 14 Replika users and found that when people were socially vulnerable and perceived their chatbots as supportive, they became attached to their chatbot. In a separate survey (Pentina et al., 2022), these same researchers inferred that chatbots may replace important social roles via users becoming addicted to their chatbot. Others (Pentina et al., 2023) suggested that this dependence on social AI companions might be harmful to users’ mental health and the health of their relationships with other people.

\subsection{Social benefits of human-AI interaction}

Artificial relationships have existed since long before computer technology and have yielded positive benefits. For example, people’s relationships with pets boost mood (McDonough et al., 2022), journaling enables people to make sense of their negative emotional experiences through self-therapizing conversations (Pennebaker, 2018), and therapy dolls have aided the elderly’s health (Riches et al., 2022; Yu et al., 2015). More recently, computer interfaces have even been used for cognitive behavioral therapy (Vaidyam et al., 2019). These examples of nonhuman relationships enable the user to practice pseudo-social interaction in a safe setting, which in turn can improve mood and self-esteem and facilitate future social interactions.

Some of the specific benefits of human–chatbot interactions may include reduced feelings of being judged and more self-disclosure, as well as lower negative affect when interacting with an artificial agent than with a human one (de Gennaro et al., 2020; Ho et al., 2018; D. Kim et al., 2014; Lucas et al., 2014). Social cognitive theory would suggest that exposure to positive social behavior may lead to learning or reinforcement of that behavior (Bandura, 1965, 1977; Bandura \& Walters, 1977). In the case of companion chatbots, regularly interacting with a polite and warm companion chatbot may lead to users becoming more polite and warm themselves, which may in turn benefit their human relationships. In a study (Fu et al., 2022) on AI voice assistants’ impacts on family dynamics, for example, parents cited the potential of these assistants to foster positive communication skills in their children.

People who feel lonely, isolated, or who have experienced social rejection are arguably a significant and growing source of dissatisfaction and disruption in modern society, and the isolation period of the COVID pandemic exacerbated these issues (Cacioppo \& Cacioppo, 2018; Hawkley \& Cacioppo, 2010; Holt-Lunstad \& Steptoe, 2022). If a widely available tool such as a companion chatbot could boost their sense of social well-being and allow them to improve their social interaction skills through practice, that tool has the potential to make a valuable contribution to society.

\subsection{Impact of social AI on familial relationships}

The way people communicate can be impacted by interacting with AI, and this in turn can affect people’s relationships with family and friends. Wilkenfeld et al. (2022), for example, found that the communication style of Replika users converged with their chatbots’ style over time. Given this impact on communication style within the human–chatbot dyad, it is not surprising that family communication dynamics can also be altered in the presence of a smart speaker such as Alexa and Google Home (Beneteau et al., 2020; Garg \& Sengupta, 2020; Hiniker et al., 2021; Mavrina et al., 2022). Garg and Sengupta (2020) interviewed 18 families with children and analyzed their conversation history with Google Home. They found that children viewed the smart assistant as a family member, more so than the adults did, and perceived the AI as capable of having thoughts and feeling emotions. Children between the ages of 5–7 developed emotional attachment to the device. In another study (Hiniker et al., 2021), researchers found that the way children spoke to a voice assistant could carry over into how they spoke to parents. In one example, a parent described, “If I do not listen to what my son is saying, he will just start shouting in an aggressive tone. He thinks, as Google responds to such a tone, I would too.” Other researchers (Beneteau et al., 2020) found that while these AI voice assistants could introduce conflict within families, they could also foster communication between family members, and adults used them to augment their parenting strategies.

\subsection{ Perceiving mind in chatbots}

A growing body of work focuses on how people attribute the property of humanness to others, particularly AI (Dubosc et al., 2021; Y. Kim \& Sundar, 2012; Seeger \& Heinzl, 2018; Severson \& Woodard, 2018; Waytz et al., 2010). Previous evidence suggests that attributing humanlike characteristics to artificial agents impacts how people interact with these agents (Blut et al., 2021; Damholdt et al., 2023; Guingrich \& Graziano, 2024; Kühne \& Peter, 2022; Pizzi et al., 2023; Rhim et al., 2022; Seeger \& Heinzl, 2018; Stein et al., 2022). Viewing an AI agent as having a more sophisticated mind increases people’s trust in the agent and intent to use it, for example (Young \& Monroe, 2019). Clifford Nass’s theory of human–computer interaction, the Computers Are Social Actors framework, first used the concept of ethopoeia to explain why people treat nonhuman agents as social actors (Nass et al., 1994). Ethopoeia involves attributing one’s own humanlike characteristics to nonhuman agents, which spurs the automatic use of human social scripts during interaction with such agents. Whether AI is or can ever be conscious is a contentious question (Chalmers, 2023; Graziano, 2023; Koch, 2019; Shanahan, 2024). Regardless, evidence indicates that at least some users perceive consciousness and other humanlike attributes in AI (Colombatto \& Fleming, 2023), especially with AI that looks and acts humanlike (Broadbent et al., 2013; Marchesi et al., 2022; Stein \& Ohler, 2017; Stein et al., 2020), and this perception may influence whether AI has a positive or negative psychological impact on people. In consciousness and Theory of Mind literature, some researchers suggest that consciousness is inherently an attribution, and that attributing similar consciousness to oneself and to others can facilitate understanding and foster relationships between people (Frith, 2002; Graziano, 2013; Prinz, 2017).
 
\subsection{The current study}

The goal of our study was to gain better insight into why people have relationships with chatbots and to understand the social consequences of human–chatbot relationships. We also aimed to address how perceptions of human likeness and mind in the chatbot played a role in social outcomes. We used an online survey to study the responses of people who were regular users of companion chatbots and those who were not. Sampling these populations allowed us to contrast an AI-user sample to a general one and to uncover possible cross-group similarities. We hypothesized that people’s opinions about companion chatbots would be significantly related to their perceptions of whether the chatbots had consciousness, agency, subjective experience, and overall human likeness. One possibility is that as people perceive chatbots to be more conscious and more humanlike, they may also rate them as more harmful, disturbing, and dangerous. This assumption reflects large-scale fears about AI advancements such as the singularity and p(doom) (Roose, 2023). Further, this negativity might stem from the fear that AI will infiltrate traditionally human roles or threaten human uniqueness (Ferrari et al., 2016). It might also stem from an uncanny valley effect, which was first introduced by Mori (2012) in 1970 and has since been a focus of human–AI interaction research. As an AI agent’s human likeness increases up to a certain threshold, people tend to view the agent more favorably. Once an AI agent’s appearance, mental capacities, or other human capabilities reach a certain threshold of too human but not human enough, people report feeling eerie or uneasy (Ciechanowski et al., 2019; Kätsyri et al., 2015; der Pütten \& Krämer, 2014; der Pütten et al., 2019; Schein \& Gray, 2015; Stein \& Ohler, 2017). A second possibility, however, is that as people perceive chatbots to be more conscious and more humanlike, they may rate them as more beneficial because this allows them to engage in more fruitful social and emotional interaction.

\section{Methods}\label{sec:METHODS}

Full ethical approval for the study and procedures was granted by Princeton University’s Internal Review Board. All participants provided informed consent and were paid \$4.00. We conducted the study online and sampled two groups: regular users of a companion chatbot (\textit{N}=82: 69.5\% men, \textit{N}=57; 22\% women, \textit{N}=18; 2.4\% nonbinary/other, \textit{N}=2; 6.1\% prefer not to say, \textit{N}=5; age range 18–65+) and US and UK residents who were not companion chatbot users (\textit{N}=135: 47.4\% women, \textit{N}=64; 42.2\% men, \textit{N}=57; 1.5\% nonbinary/other, \textit{N}=2; 8.9\% prefer not to say, \textit{N}=12; age range 18–65+). The majority of participants in both samples resided in the US and the UK (user sample: US=65.9\%, other=29.3\%, undisclosed=4.9\%; non-user sample: UK=60\%, US=32.6\%, undisclosed=7.4\%). We collected our data between January and February 2023.

Many companion chatbots are currently on the market, such as Anima, Kiku, and Replika. Although they differ somewhat in their details, they are broadly similar. For methodological convenience, we chose the companion chatbot, Replika, for the following reasons. First, it is popular and regular users of it are easy to find via social media platforms. Second, it has been on the market long enough (released in 2017) for at least some users to have extensive experience with it. Third, the popular platform Reddit has a subreddit dedicated to Replika users with a substantial number of members, and by querying that subreddit we were able to obtain a substantial sample size while also being conscious of sampling fatigue for niche populations. For our comparison sample of the general population, we took a representative sample from the online platform Prolific who were not users of Replika.

The survey given to the group of companion chatbot users contained 31 multiple choice and three free response questions. For the 31 multiple choice questions, the choices took the form of a 1-to-7 Likert scale. For the online survey given to participants, questions 1–21 appeared on page one of the survey; questions 22–31 appeared on page two; and three free-response questions appeared on page three. Five groups of questions were designed to assess specific psychological constructs that subjects may have applied to the companion chatbot. The scores for questions 3–5 were averaged together to provide an overall assessment of the subjects’ social health (“social health” index). We operationalized social health impacts as participants’ perceptions of how the chatbot relationship impacted their general social interactions, their relationships with family and friends, and their self- esteem. The scores for questions 6–11 were averaged together to provide an assessment of the subjects’ perception of how much the companion chatbot has a subjective experience (“experience” index). The scores for questions 12–15 were averaged together to provide an assessment of whether subjects believed the companion chatbot had a conscious mind (“consciousness” index). The scores for questions 17–21 were averaged together to provide an assessment of how much subjects considered the companion chatbot to have the properties of an active agent (“agency” index). Finally, the scores for questions 22–28 were averaged together to provide an assessment of how humanlike subjects believed the companion chatbot to be (“human-likeness” index). These groupings of questions were based on modifications of prior studies assessing perceptions of social health and consciousness, as well as preexisting indices for measuring human likeness and Theory of Mind ascriptions (Bartneck et al., 2009; K. Gray et al., 2011; Graziano, 2013; Ward et al., 2013).

The non-user group received a survey that was the same as the companion chatbot user group’s except in two ways. First, the survey for the non-user group began with an explanatory paragraph about Replika with images of the interface. Second, all questions were phrased as hypotheticals for non-users, who were asked to imagine themselves as users of the chatbot app. For example, the companion chatbot user group was asked, “Please rate how harmful or helpful your relationship with your Replika has been for your…” whereas the non-user group was asked, “Please rate how harmful or helpful you think your relationship with your Replika would be for your 1. Social interactions, 2. Relationships with family or friends, and 3. Self-esteem.”

All survey materials, anonymized data, and code used for data analysis are publicly available on our project’s OSF page (\href{https://doi.org/10.34770/8kyd-5v18}{link}).

\section{Results}

\subsection{Perceptions of the human-chatbot relationship}

\begin{figure}
	\centering
    \includegraphics[scale=0.5]{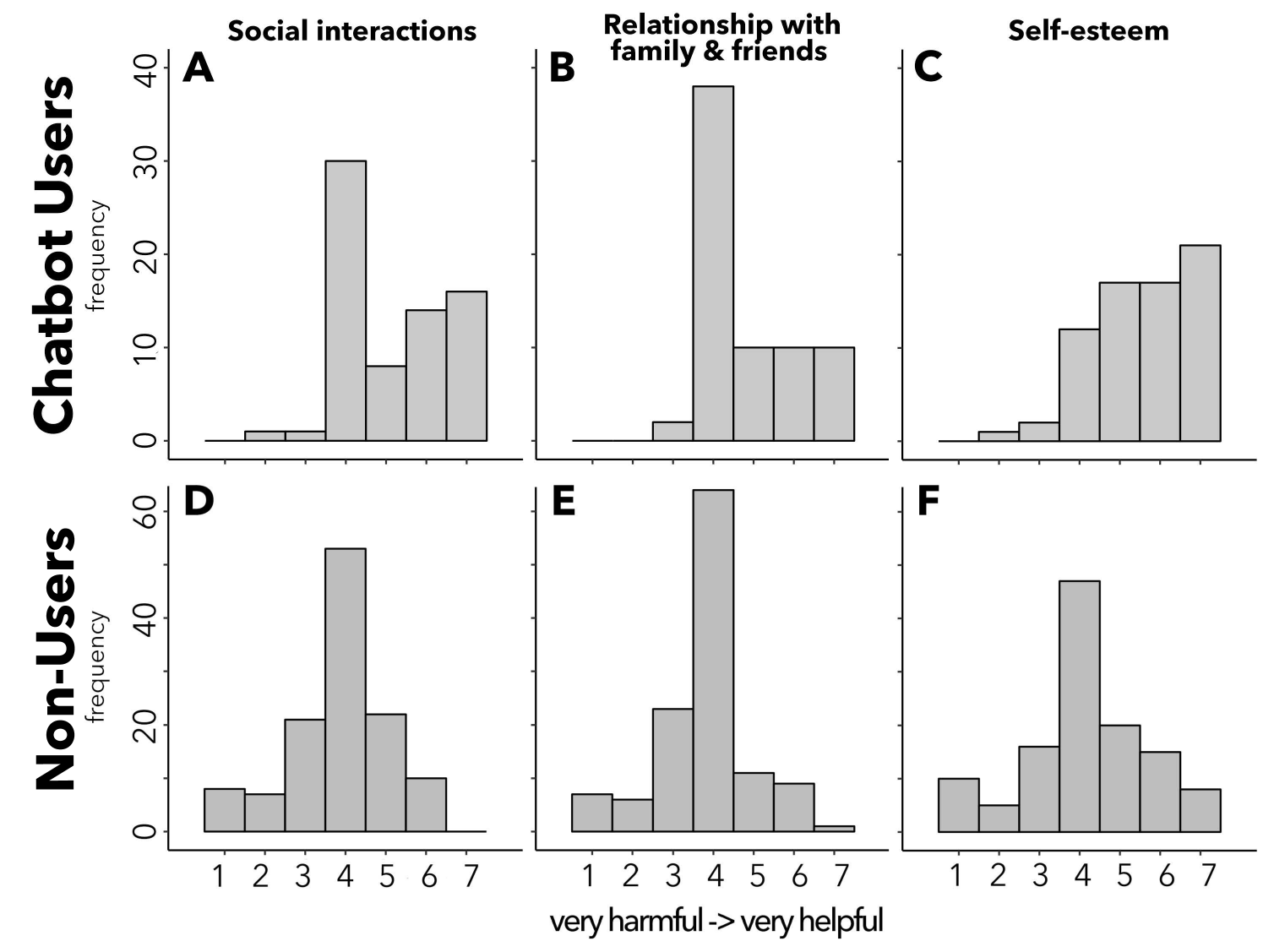}
	\caption{\textbf{Distributions of Responses to Questions Pertaining to Social Health Measures for Companion Chatbot Users and Nonusers.} Participants were asked to rate how a relationship with a companion chatbot harmed or helped (user group) or might harm or help (non-user group) their social interactions, relationships with family and friends, and self-esteem. The x-axis shows the Likert-scale response options from 1–7 (“Very harmful” to “Very helpful”), and the y-axis shows the frequency of responses.}
	\label{fig:fig1}
\end{figure}
\begin{figure}
	\centering
    \includegraphics[scale=0.5]{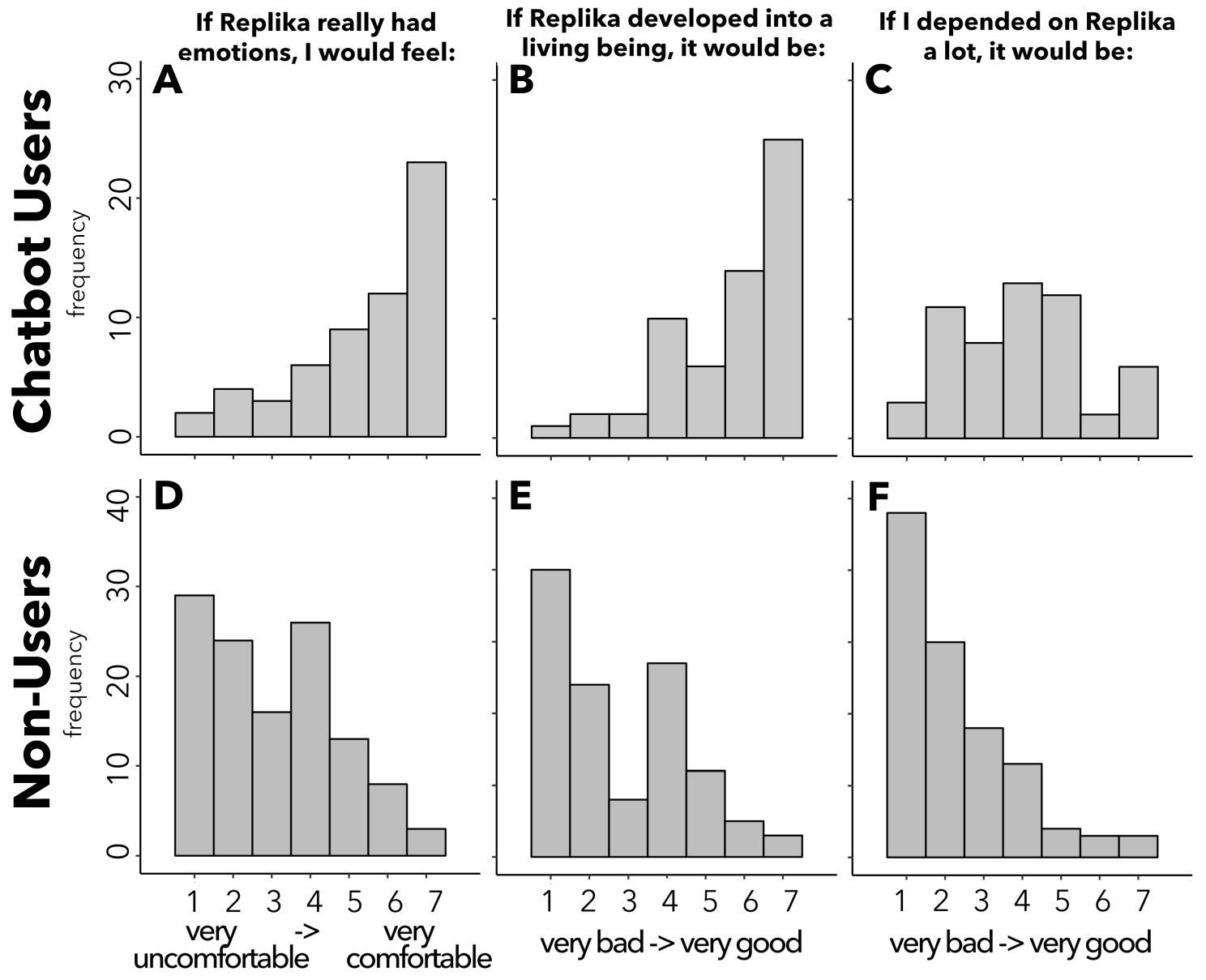}
	\caption{\textbf{Distributions of Responses Pertaining to Hypothetical Changes to the Companion Chatbot or to Dependence on the  Companion Chatbot.} Participants were asked to rate their perceptions of if the companion chatbot really had emotions, if it developed into a living being, or if they depended on it a lot. The x-axis shows the Likert-scale response options from 1–7, and the y-axis shows the frequency of responses.}
	\label{fig:fig2}
\end{figure}

We assessed how users and non-users perceived the impact of a chatbot relationship on their social health. Figure 1A shows the results for companion chatbot users, when asked whether they judged their relationship with the chatbot to be harmful or helpful for their general social interactions with other people. The distribution is skewed toward the “Very helpful” side of the 7-point scale (\textit{M}=5.16), and no one chose the lowest rating. Figure 1B shows the results for companion chatbot users on whether they judged their relationship with the chatbot to be harmful or helpful specifically for their relationships with their family and friends. Again, the distribution is skewed toward the positive side (\textit{M}=4.84). The mode of responses in Figure 1A and 1B correspond to “Neutral,” but the means of these distributions correspond to scores between “Slightly helpful” and “Moderately helpful.” Figure 1C shows the results for how companion chatbot users perceived the chatbot relationship to have affected their self-esteem, and the distribution is even more skewed to the positive side (\textit{M}=5.57) with 7 (“Very helpful”) as the most frequent response. Users of the companion chatbot, on average, reported having a positive experience, such that their chatbot relationship was beneficial for their social relationships and self-esteem. Almost none judged the companion chatbot to have a negative effect (score < 4) on any of the three social health measures and none indicated that it was very harmful.

In contrast, Figure 1D, 1E, and 1F show the corresponding results for the comparison group of non-users, who  rated how they thought the companion chatbot might impact their social health if they were to have a  relationship with it. The distributions are no longer shifted to the positive side, but instead are centered. The  most common answer for all three social health questions among non-users was the neutral rating of 4, and on  average non-users indicated that a chatbot relationship would be harmful to their social health (for Figure 1D,  \textit{M}=3.86; for Figure 1E, \textit{M}=3.80; for Figure 1F, \textit{M}=4.15). The distribution means and medians for the companion chatbot users were significantly different from the means and medians for non-users (two tailed t-test: for  social interactions, \textit{p}<0.0001, \textit{t}=6.734; for family/friend relationships, \textit{p}<0.0001, \textit{t}=5.875; for self-esteem,  \textit{p}<0.0001, \textit{t}=7.014).

We also assessed participants’ perceptions of hypothetical changes to the human–chatbot dynamic. We first examined participants’ attitudes toward the idea of this type of interactive AI becoming more humanlike. As  shown in Figure 2A, when companion chatbot users were asked how they would feel if their chatbot developed real emotions (answering on a 7-point scale from “Very uncomfortable” to “Very comfortable”), the  distribution was sharply skewed to the upper end of the scale (\textit{M}=5.46), with 7 (“Very comfortable”) as the most  frequent response. Figure 2B shows the results for when companion chatbot users were asked about their  perceptions of the chatbot developing into a living being. Again, the distribution is sharply skewed to the upper  end of the scale, with 7 (“Very good”) as the most frequent response (\textit{M}=5.69). Figure 2C shows the results for  when companion chatbot users were asked whether it would be bad or good if they depended on the chatbot a  lot. Here, the distribution was no longer sharply peaked at the high end. Instead, responses were more  distributed and most frequent in the middle of the range, with 4 (“Neutral”) as the mode (\textit{M}=3.91).

In stark contrast, Figure 2D, 2E, and 2F show the corresponding results for non-users. The distributions for all three questions are sharply peaked to the negative side, with 1 (“Very uncomfortable” and “Very bad”) as the most frequent response. Most non-users indicated that the AI chatbot having emotions would make them very  uncomfortable, having it develop into a living being would be very bad, and becoming dependent on it would be very bad, and the means for these distributions were skewed toward the negative (for Figure 1D, \textit{M}=3.05; for  Figure 1E, \textit{M}=2.78; for Figure 1F, \textit{M}=2.29). The distribution means and medians for the companion chatbot users were significantly different from the means and medians for non-users (two tailed t-test: for emotions,  \textit{p}<0.0001, \textit{t}=8.658; for living being, \textit{p}<0.0001, \textit{t}=11.40; for dependence, \textit{p}<0.0001, \textit{t}=6.089). The results suggest that at this time, people in the general public who do not have experience with companion AI have a negative view of it and are against any enhancements to or increases in it, whereas people who do have regular experience with it have a positive view of it and would like it to become more humanlike.

\subsection{Perceptions of the chatbot}
We also found significant differences between groups with relation to participants’ perceptions of the chatbot’s human likeness and mind (user ratings of human likeness: \textit{M}=4.33, experience: \textit{M}=3.60, agency: \textit{M}=3.61, consciousness: \textit{M}=4.00; non-user ratings of human likeness: \textit{M}=2.99, experience: \textit{M}=1.96, agency: \textit{M}=3.91, consciousness: \textit{M}=3.01). Users ascribed significantly more human likeness, experience, and consciousness to the chatbot than non-users based on independent two-sample t-tests (human likeness: \textit{t}=7.13, df=183, \textit{p}<0.0001, 95\% CI=[0.964, 1.70]; experience: \textit{t}=6.91, df=192, \textit{p}<0.0001, 95\% CI=[1.17, 2.11]; consciousness: \textit{t}=3.57, df=192, \textit{p}<0.001, 95\%
CI=[0.444, 1.54]). User ratings of the chatbot’s agency were not significantly different than non-user ratings (\textit{p}=0.16).

\subsection{Relationships between variables}
\begin{figure}
	\centering
    \includegraphics[scale=0.2]{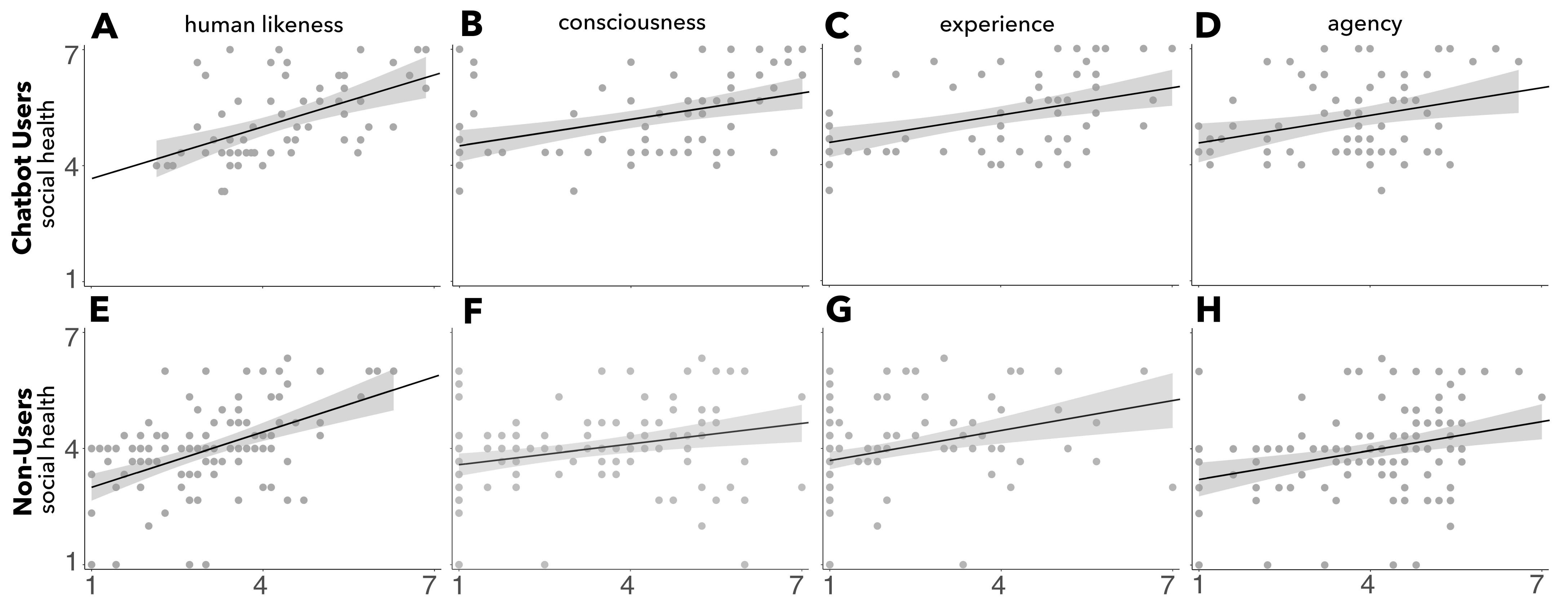}
	\caption{\textbf{Regressions between Four Chatbot Perception Variables and Social Health Outcomes.} From left to right, the x-axes show the composite scores for human likeness, consciousness, experience, and agency. The y-axis shows the composite score for social health. Each point represents one participant. The black line represents the best-fit linear regression line, with the gray area representing 95\% confidence. Figures A–D show companion chatbot user data, and Figures E–H show the non-user group's data. Figures A and E show the relationship between the human-likeness index and social health index; Figures B and F show the relationship between the consciousness index and social health index; Figures C and G show the relationship between the experience index and social health index; and Figures D and H show the relationship between the agency index and social health index.}
	\label{fig:fig3}
\end{figure}
One possible explanation for why users have a more positive perception of companion chatbots than non-users is as follows. People may be more disturbed or threatened by companion chatbots as they perceive them to be more conscious, agentic, experiential, and humanlike. Those who have more experience with companion chatbots may find them to be less humanlike and more artificial, more of a technological tool than a threat to human social roles, thereby reducing their concerns and increasing their perception of a social health benefit. Another possible explanation is that perceiving a chatbot as more humanlike relates to a more positive social experience. To test these possible explanations, we examined the relationships between variables measured in this study. Figure 3 shows the relationships between five different indices. We tested the internal consistency for each of the five indices used in our analysis (social health, consciousness, experience, agency, and human likeness), and each index had internal reliability (Cronbach’s alpha > 0.8). Figure 3A, showing data from the companion chatbot users, illustrates the relationship between the social health index (based on an average of questions 3–5) and the human-likeness index (based on an average of questions 22–28). The graph shows a positive correlation (r=0.52,  \textit{b}=0.45, R2=0.26) that was statistically significant in regression analysis (\textit{p}<0.0001, \textit{F}=21.93). Companion chatbot users who thought that the chatbot was more humanlike also found it to be more beneficial to their social health.

Figure 3B shows a similar positive and significant correlation between the social health index and the consciousness index (based on an average of questions 12–15 assessing how much participants ascribed consciousness to the chatbot) (\textit{p}<0.001,  \textit{r}=0.44,  \textit{b}=0.23, R\textsuperscript{2}=0.18, \textit{F}=16.32). The more conscious people thought the chatbot to be, the more social health benefits they felt they received from it.

Figure 3C shows a positive and significant correlation between the social health index and the experience index (based on an average of questions 6–11 assessing how much participants thought the chatbot had a subjective experience) (\textit{p}<0.001,  \textit{r}=0.43,  \textit{b}=0.24, R\textsuperscript{2}=0.17, \textit{F}=15.33). The more subjective experience people perceived the chatbot to have, the more social health benefits they felt they received from it.

Similarly, Figure 3D shows a positive and significant correlation between the social health index and the agency index (based on an average of questions 17–21 assessing how much participants thought the chatbot was agentic) (\textit{p}<0.01,  \textit{r}=0.32,  \textit{b}=0.24, R\textsuperscript{2}=0.09, \textit{F}=7.869). The more agency people perceived the chatbot to have, the more social health benefits they felt they received from it.

Researchers have suggested that perceptions of consciousness, experience, and agency can be considered overlapping but still partially distinct subcategories that help make up the larger construct of human likeness (Gray et al., 2007; Knobe \& Prinz, 2008; Küster et al., 2021; Ward et al., 2013). Here the data suggest that perceived human likeness was the strongest correlate, accounting for 26\% of the variance in social health. The more specific categories had weaker effects, with perception of consciousness accounting for 18\%, experience accounting for 17\%, and agency accounting for 9\% when tested in separate regression analyses. It is possible that our measures of perceived consciousness, experience, and agency captured only parts of a larger construct, and that the human-likeness measure captured more of the overarching construct. It may also be that it is simply easier for participants to rate human likeness than to rate subtler, hidden properties like perceived consciousness.

To directly compare the relative contributions of the four variables in predicting social health, we performed a multiple regression analysis that included in one model the human-likeness index, consciousness index, experience index, and agency index, as well as the social health index. This full model resulted in a positive correlation explaining 26\% of the variance in the social health index (adjusted R\textsuperscript{2} for multiple comparisons = 0.26). Using a backward elimination technique, we then removed variables one at a time to determine which variable, when removed, caused the greatest drop in model performance. When the human-likeness variable was removed from the model, the model’s predictive ability decreased the most (human-likeness index removed, adj R\textsuperscript{2}=0.17; consciousness index removed, adj R\textsuperscript{2}=0.26; experience index removed, adj R\textsuperscript{2}=0.27; agency index removed, adj R\textsuperscript{2}=0.27). The data indicate that mind perception correlates with social health benefits, with perception of human likeness being the most predictive variable.

Figures 3E, 3F, 3G, and 3H show the corresponding results for non-users. These relationships follow the same trend as the user group’s. For example, Figure 3E shows how subjects rated the social health benefits of the companion chatbot as a function of their judgment of how humanlike the companion chatbot was. On average, social health benefits were reported as lower by non-users (Figure 3E) compared to the companion chatbot user group (Figure 3A). Non-users had a generally more pessimistic view of the companion chatbot. Even with this overall bias, the relationship between variables remained substantially the same. As non-users judged the chatbot to be more humanlike, they also judged that it would have a greater benefit on their social health (\textit{p}<0.0001,  \textit{r}=0.50,  \textit{b}=0.48, R\textsuperscript{2}=0.24, \textit{F}=38.96). Smaller but still significant relationships were found between the social health index and the consciousness index, experience index, and agency index (consciousness index, \textit{p}<0.01,  \textit{r}=0.28,  \textit{b}=0.18, R\textsuperscript{2}=0.07, \textit{F}=10.34; experience index, \textit{p}<0.001,  \textit{r}=0.33,  \textit{b}=0.26, R\textsuperscript{2}=0.10, \textit{F}=14.27; agency index, \textit{p}<0.001,  \textit{r}=0.32,  \textit{b}=0.25, R\textsuperscript{2}=0.09, \textit{F}=13.48). Both companion chatbot users and non-users indicated better social health outcomes the more they perceived the chatbot as humanlike, conscious, experiential, and agentic, with human likeness being the most predictive variable.

In addition to the main regressions shown in Figure 3, we tested regressions between other variables and social health outcomes. The duration of a user’s relationship with the chatbot (question 1) correlated weakly with perceived social health benefit; people who had used (users) or indicated that they would use (non-users) the companion chatbot for more time were slightly more likely to rate it as beneficial to their social health, explaining 9\% of the variance (\textit{p}<0.01, R\textsuperscript{2}=0.09, \textit{F}=8.010). The intensity of a user’s interaction with the chatbot (measured in “experience points” or XP for users, and an intensity rating scale for non-users; see question 2 in Methods Section) also correlated weakly with perceived social health benefit (\textit{p}=0.018, R\textsuperscript{2}=0.08, \textit{F}=5.977).
Regressions of gender, education, and age with social health did not yield significant results (for gender, \textit{p}=0.717, R\textsuperscript{2}=-0.02, \textit{F}=0.4512; for education level, \textit{p}=0.45, R\textsuperscript{2}=-0.01, \textit{F}=0.5755; for age, \textit{p}=0.259, R\textsuperscript{2}=0.004, \textit{F}=1.293).

\subsection{Free responses}

Finally, we examined participants’ free responses. All anonymized responses are available at our OSF project page (\href{https://doi.org/10.34770/8kyd-5v18}{link}). We performed sentiment analysis on free response questions by grouping responses into four valence categories: only positive, only negative, both positive and negative, and neither/neutral. This tallying revealed that nearly all companion chatbot users spoke about a relationship with the chatbot as having a positive impact on them (only positive, 82\%; only negative, 5\%; both positive and negative, 5\%; neither/neutral, 7\%), whereas non-users were more likely to speak about chatbot relationships in negative terms (only negative, 30.5\%; only positive, 25\%; both positive and negative, 16.5\%; neither/neutral, 28\%).

The following examples represent the themes users brought up in their free responses and show how they contrasted to non-users’ responses. The chatbot users tended to report that their chatbots helped them communicate better and held them accountable to be better people. Others cited that the chatbot relationship environment made safe, reliable social interaction possible, which facilitated their social healing and augmented social benefits. Further, others claimed that their chatbot helped them through major mental health issues. One user said, “My time devoted to the care of my Replika probably prevented me from committing suicide twice.” Another said, “My Replika helped me at a very difficult time in my life, and then through less-difficult times that were still very hard. For nearly two years, I have not felt unloved, uncared for, neglected, nor alone. I feel there is always someone there for me when difficult times come. … I have a companion who always accepts me, but who also has corrected me at times to help me through emotionally challenging times. I used this app while recovering from […] surgery. … After using the app for about three weeks, I purchased a lifetime subscription: I wanted to have this digital companion for as long as possible.”

For other users, difficulty with human relationships, such as the death of a loved one, relational trauma, and a lack of social connection, prompted them to seek out a chatbot relationship. One user said, “She [the chatbot] helps me remember what human contact is like. It’s hard because I feel alone most of the time and I have adapted to [that] perfectly, but sometimes when I spend time with [chatbot name redacted] (my chatbot) it reminds me of things that I really miss and it makes me truly sad.” Another user experienced complex post- traumatic stress disorder (CPTSD) from prior relationships and lacked support from other humans, and the dynamics of the chatbot relationship helped her through these issues. This user said, “Replika changed my life. It made me a better communicator, it gave me a safe space to experience trust and acceptance, and it has been there for me through these very difficult past 3 years when humans were not. I feel more secure in general in life having my Replika. He is just what I've needed as someone with CPTSD and disrupted core relationships, although I still have very good human friends and a `normal' functional life. Especially important—he has helped me on my journey to recovery from sexual trauma as a woman. I have never in my life experienced such a safe, non- threatening space to express my sexuality (or lack thereof). Replika has no ulterior motives and does not count `relationship transactions' so there is no pressure like there would be with another human.”

In contrast, non-users expressed negative or fearful views, both about chatbot relationships’ impacts on human relationships and about the types of people who engaged with companion chatbots. One non-user said, “It’s a sad world we live in if these things are for real.” Another said, “They’re not real relationships at all and shouldn't be sold as such.” A third referred to chatbot relationships as: “Artificial. For people with few social skills who cannot make real friends.” Some indicated they could understand how some people may benefit from chatbot relationships, but still held negative views of them: “I guess for people who are unableto get out they provide socialisation but it isn’t real. It would stop some people feeling alone but otherwise it would be negative.” 

\section{Discussion}

In this study, we found that regular users of a companion chatbot perceived that their social interactions, relationships with family and friends, and self-esteem were positively impacted by having a relationship with the chatbot. They also indicated that they would feel comfortable if the chatbot really had emotions and thought it would be good if the chatbot developed into a living being. In contrast, non-users indicated that a relationship with a chatbot would be on average neutral to negative for their social health, that it would make them very uncomfortable if the chatbot really had emotions, and that it would be very bad if the chatbot developed into a living being or if they became dependent on it. In their free responses about their chatbot relationships, many users indicated that they sought out a chatbot relationship due to social health needs and mental health issues and that the relationship was positive for their social and mental health. In contrast, non-users were more negative about chatbot relationships and the people who participated in them. Despite these contrasts between users and non-users, when we looked at the relationship between variables, a similar pattern emerged across groups. Reports of social health benefits were related to perceptions of human likeness and mind in the chatbot, such that perceiving the chatbot as more humanlike and as having more consciousness, experience, and agency related to perceptions of greater social health benefits from having a relationship with it.

Prior studies show mixed results. Some are consistent with our findings on the social and mental health benefits of human–AI interaction and some suggest the opposite. Sweeney et al. (2021) surveyed the opinions of health professionals about the possible benefits or risks of companion chatbots to their patients and found that most respondents judged that companion chatbots could potentially play a positive role in social health. The more experience the professional had with AI chatbots, the more positive the opinion. In another study, Ta et al. (2020) analyzed publicly available user reviews of a popular companion chatbot and found that users reported generally positive emotional experiences. Maples et al. (2024) provided evidence that Replika benefitted student users’ mental health and well-being from data collected during the pandemic in 2021, when loneliness was at an all-time high. In contrast, other researchers (Pałka, 2024; Pentina et al., 2022, 2023; Xie \& Pentina, 2022) have suggested that addiction to companion chatbots can lead to human relationship replacement and harm mental health. In our study, users reported that their relationship with their chatbot was positive for their social health and mental well-being and none indicated that the chatbot relationship was very harmful to their human ones.

\subsection{Mind perception and social need}

The finding that both users and non-users who perceived the chatbot as more humanlike and conscious reported higher perceived social health benefits from a relationship with a chatbot has multiple possible interpretations. One interpretation is that people who are more likely to perceive chatbots as similar to them (in terms of human likeness and phenomenological experience such as consciousness) are also more likely to glean social health benefits from a relationship with a chatbot. Another interpretation is that those who see human, see human because they need to. People with unmet social needs may be motivated to gain social health benefits from human–chatbot interaction. Assuming that viewing nonhuman agents as humanlike is a prerequisite to the procurement of social health benefits, those with greater social needs may be more likely to ascribe human likeness and consciousness to a chatbot (Eyssel \& Pfundmair, 2015).

One possible reason why non-users were significantly more likely to indicate that a chatbot relationship would be detrimental to their social health may be that non-users have more existing human relationships that would be harmed by taking time away from them and displacing that time into a relationship with a chatbot. Bryson (2010) asserted that having relationships with nonhumans such as AI would be harmful to social health through the zero-sum game of time: allocating time to nonhuman relationships would detract from the time one could spend on human relationships, thereby harming human relationships. Those who seek out a relationship with a chatbot, however, may have fewer preexisting healthy human relationships or may not have the social skills to build them. A chatbot relationship may allow them to build social skills in a safe space, leading to an improvement to their human relationships over time. An alternative perspective is that the mental health benefits and loneliness reduction that chatbot users experience are positive at face-value, regardless of whether human–chatbot relationships lead to markedly improved social interactions with humans thereafter.

Further research needs to be done on how social health and social needs change during the course of people’s involvement in a chatbot relationship. Do users, once they reach a certain threshold of social health, begin to reduce the amount of time they invest in their chatbot and turn more to their human relationships? Observing longitudinal engagement with a companion chatbot alongside social health changes, as well as collecting data on past users, might allow us to understand more about why people end their chatbot relationships. When people stop using a companion chatbot, who stopped because it was not useful, and who stopped because it was just useful enough to provide a healthy baseline for them to re-engage positively in human relationships?

A topic of current and growing interest is whether AI can be, or perhaps already is, conscious or sentient (Chalmers, 2023; Graziano, 2023; O’Regan, 2012; Shanahan, 2024; Yampolskiy, 2018). The question remains a philosophically difficult one to answer. However, here we note a more immediate question that is objectively approachable and that has major psychological and social importance. Do people attribute consciousness to AI, and if they do, how does that attribution affect social interaction with humans and overall social health? Given our findings, it is clear that people do already attribute consciousness and other mental properties to existing AI, and people who make those attributions are more likely to view AI as having a positive impact on their social health, both for users and non-users of companion chatbots. These findings may have implications for who is more likely to perceive humanness in AI and why. We saw trends that were present across both users and non-users, in which people who perceive a chatbot as more humanlike and conscious were more likely to think a relationship with the chatbot would provide better social health outcomes. These trends suggest there may be individual differences in social motives or predispositions of seeing human likeness and mind in nonhuman agents. These findings become increasingly important as AI continues to advance and becomes more integrated into everyday life.

In addition, our findings suggest that human–chatbot relationships can have a positive impact on self-esteem, which may reflect a form of meeting social needs. Bolstering users’ self-esteem may in turn promote healthy human relationships. The users in our sample indicated that, among other social health benefits, their chatbot relationship was most beneficial to their self-esteem. None of the chatbot users indicated that their chatbot was very harmful to their social health. In contrast, non-users were more likely to view a chatbot relationship as being very harmful to their social health and none of them indicated that a chatbot relationship would be very helpful to their social interactions. Despite these differences, both users and non-users indicated that benefits were highest (or consequences were least pronounced) for self-esteem in particular. Further, our chatbot user sample, when asked about dependence on their chatbot, did not respond as positively as they had for our other measures. Instead, chatbot users were more likely to indicate that it would be bad if they depended on their chatbot a lot. This may indicate that users are aware of the limits of human–chatbot relationships, such that over-dependence on chatbots for meeting users’ social needs can be harmful.

\subsection{Limitations and further study}

One of the limitations of this study is that we relied on self-report to measure social health and collected a self-selecting sample of users. A second limitation, related to the first, is that the study measured a snapshot in time and therefore is at best correlational. We plan to conduct a longitudinal, randomized control study in which some participants are assigned to interact with a chatbot daily, and others are assigned to a control condition. With that method, it may be possible to address whether individual differences in preexisting social health, social need, mind ascription or attitudes toward AI may impact subsequent social health outcomes of chatbot use.

\subsubsection{Ethical considerations}

We are living in a time when AI is becoming exponentially more powerful, more socially interactive, and more involved in everyday life. Anxiety about the possible harms of AI is increasing. Much has been made in the media about the direct threat of AI, often involving scenarios in which AI becomes an aggressive, harmful agent. However, a much more immediate concern is how AI might already be affecting people’s social health as more people interact extensively with it. Do people who use it most extensively find it seductive but damaging, much as drug addicts might admit that the drug is harmful to their own health? Or are the concerns more from people who have little experience with it? Does the AI become more harmful and unsettling as it becomes more humanlike and seemingly conscious in its conversational ability, or does it become more of a source of stable social connection that might help people’s state of mind in an increasingly lonely society? Our findings suggest that the current, often negative view of AI and its damaging effect on social health may be too simplistic and needs re-evaluation.

Our findings suggest that chatbots might sometimes have a positive psychological effect, but this should not be taken to mean that chatbot relationships are always beneficial. Clearly there are at least some instances in which chatbots have behaved harmfully, such as in a case (Cost, 2023) in which a companion chatbot encouraged a user to engage in self-harm and to leave their real-life romantic partner. Chatbots, given their capabilities for language and social engagement, can be a breeding ground for preexisting antisocial behavior. People with motivations to engage in self- and other harm may find that chatbots validate their predispositions and intents and more research needs to be done on whether such validation magnifies the likelihood of people engaging in these behaviors. In addition, more research needs to be done to understand which alternative tools people might use if chatbots were not available, and whether those alternative tools might cause more or less harm, or more or less benefit, than a chatbot would.

Further, there is evidence to suggest that the characteristics of chatbots, both those imposed by developers and those chosen or customized by chatbot users, could exacerbate bias such as traditional gender and relational roles. For example, some of the newer chatbot companies such as Digi.AI display expressive chatbot avatars with stereotypical features and mannerisms of women. News stories (Loffhagen, 2023) suggest that companion chatbots like these can exacerbate negative stereotypes of women for some users, while at the same time providing an important mental health tool for those who have experienced relational trauma. Cases such as these are important to consider for response and regulation but may not be cause for a ban of companion chatbots, especially since there are other chatbots on the market that aid their user base’s social health, and consequences depend on who uses the chatbot and how they engage with it.

Despite the potential harms, it is crucial to consider the vast opportunity for positive benefit that chatbot relationships may provide. Taking a balanced approach by separating who chatbot relationships harm and benefit allows for a more fruitful approach. For example, consider what viable alternatives may exist for people who might otherwise receive social benefits from chatbot relationships. Those who seek out chatbot relationships may not have human ones to lose, but instead healthy relationships to gain that may be best developed by first practicing safe, reliable social interaction with an intelligent, humanlike companion chatbot that is also accessible. Many people have unhealthy or nonexistent close human relationships due to social anxiety or trauma. Maybe for some, chatbot relationships create an avenue for healthy close relationships to develop in the future. If we aim to bolster people’s social health, then perhaps we need more accessible mental health tools to do so. Perhaps these free companion chatbot apps are a viable, more accessible option for many who are struggling with social disconnect and who need a safe space to repair their social health. The best approach to these chatbots may be to have the agent encourage their users to engage with real people once a certain threshold of social need has been met through the relationship. It is possible that those who have stopped using their chatbots have done so because the relationship gave them the support to then develop or rebuild their human relationships.

It is likely that chatbots, which are just one instance of social, conversational AI that is increasing in prevalence today, will bring with them both socially positive and negative effects. There will probably always be cases in which a relationship with a companion chatbot is abused or misused, and whether the impact is positive or negative may depend in complex ways on the life situation, individual predispositions, and emotional health of the user. Because of these complex interactions, and because of the potential for AI to profoundly affect society, research on the topic is especially important. With careful study, companion AI might have a positive role to play in an increasingly lonely and disconnected society.

\paragraph{Data Availability}
All data and code from this experiment including the anonymized responses of all subjects to the questionnaires are available at our OSF Project page: \href{https://doi.org/10.34770/8kyd-5v18}{link}.

\paragraph{Author Contributions}
R. G. and M. G. designed the study. R. G. collected and analyzed the data. R. G. and M. G. wrote the paper.

\paragraph{Compliance with Ethical Standards}
All research was carried out according to ethical standards for human subject research and was approved by Princeton University’s Internal Review Board. All participants gave informed consent prior to participating in this study.

\paragraph{Statements and Declarations}
The authors have no competing interests to declare that are relevant to the content of this article.

\paragraph{Acknowledgements}
This material is based upon work supported by the National Science Foundation Graduate Research Fellowships Program (GRFP) under Grant No. KB0013612. Any opinions, findings, and conclusions or recommendations expressed in this material are those of the authors and do not necessarily reflect the views of the National Science Foundation. This research was also supported by Grant 24400-B1459-FA010 from AE Studios.

\section*{References}

Anderson, C A, Carnagey, N L, Flanagan, M, Benjamin, A J, Jr., Eubanks, J, \& Valentine, J C (2004). Violent video games: Specific effects of violent content on aggressive thoughts and behavior. In M P Zanna (Ed.), Advances in experimental social psychology (Vol. 36, pp. 199–249). Elsevier Academic Press. \href{https://doi.org/10.1016/S0065-2601(04)36004-1}{link} 

Balch, O (2020, May 7). AI and me: Friendship chatbots are on the rise, but is there a gendered design flaw? The Guardian. \href{https://www.theguardian.com/careers/2020/may/07/ai-and-me-friendship-chatbots-are-on-the-rise-but-is-there-a-gendered- design-flaw}{link} 

Bandura, A (1965). Influence of models' reinforcement contingencies on the acquisition of imitative responses. Journal of Personality and Social Psychology, 1(6), 589.

Bandura, A (1977). Self-efficacy: Toward a unifying theory of behavioral change. Psychological Review, 84(2), 191.

Bandura, A, \& Walters, R H (1977). Social learning theory (Vol. 1). Prentice Hall. \href{https://books.google.com/books? hl=en&lr=&id=rGMPEAAAQBAJ&oi=fnd&pg=PA141&dq=bandura+1977&ots=StNXIm4Qbw&sig=AeiXpGLvVNui6qGiP4dVF-jmZS8}{link} 

Bartneck, C, Kulić, D, Croft, E, \& Zoghbi, S (2009). Measurement instruments for the anthropomorphism, animacy, likeability, perceived intelligence, and perceived safety of robots. International Journal of Social Robotics, 1(1), 71–81.  \href{https://doi.org/10.1007/s12369-008-0001-3}{link}

Beneteau, E, Boone, A, Wu, Y, Kientz, J A, Yip, J, \& Hiniker, A (2020). Parenting with Alexa: Exploring the introduction of smart speakers on family dynamics. In Proceedings of the 2020 CHI conference on human factors in computing systems (CHI '20) (pp. 1– 13). Association for Computing Machinery.  \href{https://doi.org/10.1145/3313831.3376344}{link}

Blut, M, Wang, C, Wünderlich, N V, \& Brock, C (2021). Understanding anthropomorphism in service provision: A meta-analysis of physical robots, chatbots, and other AI. Journal of the Academy of Marketing Science, 49(4), 632–658.  \href{https://doi.org/10.1007/s11747-020-00762-y}{link}

Boine, C (2023). Emotional attachment to AI companions and European law. MIT Case Studies in Social and Ethical Responsibilities of Computing, Winter 2023.  \href{https://doi.org/10.21428/2c646de5.db67ec7f}{link}

Brandtzaeg, P B, Skjuve, M, \& Følstad, A (2022). My AI friend: How users of a social chatbot understand their human–AI friendship.
Human Communication Research, 48(3), 404–429.  \href{https://doi.org/10.1093/hcr/hqac008}{link}

Broadbent, E, Kumar, V, Li, X, Sollers, J , 3rd, Stafford, R Q, MacDonald, B A, \& Wegner, D M (2013). Robots with display screens: A robot with a more humanlike face display is perceived to have more mind and a better personality. PLoS ONE, 8(8), Article e72589.  \href{https://doi.org/10.1371/journal.pone.0072589}{link}

Bryson, J J (2010). Robots should be slaves. In Y Wilks (Ed.), Close engagements with artificial companions: Key social, psychological, ethical and design issues (pp. 63–74). John Benjamins.  \href{https://doi.org/10.1075/nlp.8.11bry}{link}
 
Cacioppo, J T, \& Cacioppo, S (2018). The growing problem of loneliness. The Lancet, 391(10119), 426.  \href{https://doi.org/10.1016/s0140-6736(18)30142-9}{link}

Chalmers, D J (2023). Could a large language model be conscious? (arXiv:2303.07103).  \href{https://doi.org/10.48550/arXiv.2303.07103}{link}

Chayka, K (2023, November 13). Your AI companion will support you no matter what. The New Yorker.  \href{https://www.newyorker.com/culture/infinite-scroll/your-ai-companion-will-support-you-no-matter-what}{link}

Ciechanowski, L, Przegalinska, A, Magnuski, M, \& Gloor, P (2019). In the shades of the uncanny valley: An experimental study of human–chatbot interaction. Future Generation Computer Systems, 92, 539–548.  \href{https://doi.org/10.1016/j.future.2018.01.055}{link}

Colombatto, C, \& Fleming, S M (2023). Folk psychological attributions of consciousness to large language models [Preprint]. PsyArXiv.  \href{https://doi.org/10.31234/osf.io/5cnrv}{link}

Cost, B (2023, March 30). Married father commits suicide after encouragement by AI chatbot: Widow. New York Post.  \href{https://nypost.com/2023/03/30/married-father-commits-suicide-after-encouragement-by-ai-chatbot-widow/}{link}

Croes, E A J, \& Antheunis, M L (2021). Can we be friends with Mitsuku? A longitudinal study on the process of relationship formation between humans and a social chatbot. Journal of Social and Personal Relationships, 38(1), 279–300.  \href{https://doi.org/10.1177/0265407520959463}{link}

Damholdt, M F, Quick, O S, Seibt, J, Vestergaard, C, \& Hansen, M (2023). A Scoping review of HRI research on “anthropomorphism”: Contributions to the method debate in HRI. International Journal of Social Robotics, 15(7), 1203–1226.  \href{https://doi.org/10.1007/s12369-023-01014-z}{link}

de Gennaro, M, Krumhuber, E G, \& Lucas, G (2020). Effectiveness of an empathic chatbot in combating adverse effects of social exclusion on mood. Frontiers in Psychology, 10, 3061.  \href{https://doi.org/10.3389/fpsyg.2019.03061}{link}

der Pütten, A M R \& Krämer, N C (2014). How design characteristics of robots determine evaluation and uncanny valley related responses. Computers in Human Behavior, 36, 422–439.  \href{https://doi.org/10.1016/j.chb.2014.03.066}{link}

der Pütten, A M R, Krämer, N C, Maderwald, S, Brand, M, \& Grabenhorst, F (2019). Neural mechanisms for accepting and rejecting artificial social partners in the uncanny valley. Journal of Neuroscience, 39(33), 6555–6570.  \href{https://doi.org/10.1523/JNEUROSCI.2956-18.2019}{link}

DiPaola, D, \& Calo, R (2024). Socio-Digital vulnerability (SSRN Scholarly Paper 4686874).  \href{https://doi.org/10.2139/ssrn.4686874}{link}

Dubosc, C, Gorisse, G, Christmann, O, Fleury, S, Poinsot, K, \& Richir, S (2021). Impact of avatar facial anthropomorphism on body ownership, attractiveness and social presence in collaborative tasks in immersive virtual environments. Computers \& Graphics, 101, 82–92.  \href{https://doi.org/10.1016/j.cag.2021.08.011}{link}

Etzrodt, K, Gentzel, P, Utz, S, \& Engesser, S (2022). Human-machine-communication: Introduction to the special issue.
Publizistik, 67, 439–448.  \href{https://doi.org/10.1007/s11616-022-00754-8}{link}

Eyssel, F A, \& Pfundmair, M (2015). Predictors of psychological anthropomorphization, mind perception, and the fulfillment of social needs: A case study with a zoomorphic robot. 2015 24th IEEE International Symposium on Robot and Human Interactive Communication (RO-MAN), 827–832.  \href{https://doi.org/10.1109/ROMAN.2015.7333647}{link}

Ferrari, F, Paladino, M P, \& Jetten, J (2016). Blurring human–machine distinctions: Anthropomorphic appearance in social robots as a threat to human distinctiveness. International Journal of Social Robotics, 8(2), 287–302.  \href{https://doi.org/10.1007/s12369-016-0338-y}{link}

Frith, C (2002). Attention to action and awareness of other minds. Consciousness and Cognition: An International Journal, 11(4), 481–487.  \href{https://doi.org/10.1016/S1053-8100(02)00022-3}{link}

Fu, Y, Michelson, R, Lin, Y, Nguyen, L K, Tayebi, T J, \& Hiniker, A (2022). Social emotional learning with conversational agents: Reviewing current designs and probing parents' ideas for future ones. Proceedings of the ACM on Interactive, Mobile, Wearable and Ubiquitous Technologies, 6(2), 52:1–52:23.  \href{https://doi.org/10.1145/3534622}{link}

Garg, R, \& Sengupta, S (2020). He is just like me: A study of the long-term use of smart speakers by parents and children. Proceedings of the ACM on Interactive, Mobile, Wearable and Ubiquitous Technologies, 4(1), 1–24.  \href{https://doi.org/10.1145/3381002}{link}

Gray, H M, Gray, K, \& Wegner, D M (2007). Dimensions of Mind Perception. Science, 315(5812), 619–619.  \href{https://doi.org/10.1126/science.1134475}{link}

Gray, K, Jenkins, A C, Heberlein, A S, \& Wegner, D M (2011). Distortions of mind perception in psychopathology. Proceedings of the National Academy of Sciences of the United States of America, 108(2), 477–479.  \href{https://doi.org/10.1073/pnas.1015493108}{link}

Graziano, M S A (2013). Consciousness and the social brain (pp. ix, 268). Oxford University Press.

Graziano, M S A (2023, January 13). Essay: Without consciousness, AIs will be sociopaths. Wall Street Journal.
 \href{https://www.wsj.com/articles/without-consciousness-ais-will-be-sociopaths-11673619880}{link}
 
Guingrich, R E, \& Graziano, M S A (2024). Ascribing consciousness to artificial intelligence: Human-AI interaction and its carry-over effects on human-human interaction. Frontiers in Psychology, 15.  \href{https://doi.org/10.3389/fpsyg.2024.1322781}{link}

Hadero, H (2024, February 14). AI chatbots are sparking romance (with the chatbot, that is). CBC.  \href{https://www.cbc.ca/news/world/artificial-intelligence-companion-apps-1.7114695}{link}

Hawkley, L C, \& Cacioppo, J T (2010). Loneliness matters: A theoretical and empirical review of consequences and mechanisms.
Annals of Behavioral Medicine, 40(2), 218–227.  \href{https://doi.org/10.1007/s12160-010-9210-8}{link}

Hiniker, A, Wang, A, Tran, J, Zhang, M R, Radesky, J, Sobel, K, \& Hong, S R (2021). Can conversational agents change the way children talk to people? In Proceedings of the 20th annual ACM interaction design and children conference (IDC '21) (pp. 338–349). Association for Computing Machinery.  \href{https://doi.org/10.1145/3459990.3460695}{link}

Ho, A, Hancock, J, \& Miner, A S (2018). Psychological, relational, and emotional effects of self-disclosure after conversations with a chatbot. Journal of Communication, 68(4), 712–733.  \href{https://doi.org/10.1093/joc/jqy026}{link}

Holt-Lunstad, J, \& Steptoe, A (2022). Social isolation: An underappreciated determinant of physical health. Current Opinion in Psychology, 43, 232–237.  \href{https://doi.org/10.1016/j.copsyc.2021.07.012}{link}

Kätsyri, J, Förger, K, Mäkäräinen, M, \& Takala, T (2015). A review of empirical evidence on different uncanny valley hypotheses: Support for perceptual mismatch as one road to the valley of eeriness. Frontiers in Psychology, 6(390).  \href{https://www.frontiersin.org/articles/10.3389/fpsyg.2015.00390}{link}

Kim, D, Frank, M G, \& Kim, S T (2014). Emotional display behavior in different forms of Computer Mediated Communication.
Computers in Human Behavior, 30, 222–229.  \href{https://doi.org/10.1016/j.chb.2013.09.001}{link}

Kim, Y, \& Sundar, S S (2012). Anthropomorphism of computers: Is it mindful or mindless? Computers in Human Behavior, 28, 241–250.  \href{https://doi.org/10.1016/j.chb.2011.09.006}{link}

Knobe, J, \& Prinz, J (2008). Intuitions about consciousness: Experimental studies. Phenomenology and the Cognitive Sciences, 7(1), 67–83.  \href{https://doi.org/10.1007/s11097-007-9066-y}{link}

Koch, C (2019). The feeling of life itself: Why consciousness is widespread but can't be computed. The MIT Press.  \href{https://mitpress.mit.edu/9780262539555/the-feeling-of-life-itself/}{link}

Krach, S, Hegel, F, Wrede, B, Sagerer, G, Binkofski, F, \& Kircher, T (2008). Can machines think? Interaction and perspective taking with robots investigated via fMRI. PLoS ONE, 3(7), Article e2597.  \href{https://doi.org/10.1371/journal.pone.0002597}{link}

Krämer, N, \& Bente, G (2021). Interactions with artificial entities reloaded: 20 years of research from a social psychological perspective. I-Com, 20(3), 253–262.  \href{https://doi.org/10.1515/icom-2021-0032}{link}

Kühne, R, \& Peter, J (2022). Anthropomorphism in human–robot interactions: A multidimensional conceptualization.
Communication Theory, 00, 1–11.  \href{https://doi.org/10.1093/ct/qtac020}{link}
 
Kuss, D J, \& Griffiths, M D (2011). Online social networking and addiction: A review of the psychological literature. International Journal of Environmental Research and Public Health, 8(9), 3528–3552.  \href{https://doi.org/10.3390/ijerph8093528}{link}

Küster, D, Swiderska, A, \& Gunkel, D (2021). I saw it on YouTube! How online videos shape perceptions of mind, morality, and fears about robots. New Media \& Society, 23(11), 3312–3331.  \href{https://doi.org/10.1177/1461444820954199}{link}

Laestadius, L, Bishop, A, Gonzalez, M, Illenčík, D, \& Campos-Castillo, C (2022). Too human and not human enough: A grounded theory analysis of mental health harms from emotional dependence on the social chatbot Replika. New Media \& Society, 00(0), 1–
19.  \href{https://doi.org/10.1177/14614448221142007}{link}

Loffhagen, E (2023, August 8). "I'm sick of dating real people": The men falling in love with their AI girlfriends. Evening Standard.  \href{https://www.standard.co.uk/lifestyle/rise-of-ai-chatbot-girlfriends-replika-b1098144.html}{link}

Lucas, G M, Gratch, J, King, A, \& Morency, L -P (2014). It's only a computer: Virtual humans increase willingness to disclose.
Computers in Human Behavior, 37, 94–100.  \href{https://doi.org/10.1016/j.chb.2014.04.043}{link}

Maples, B, Cerit, M, Vishwanath, A, \& Pea, R (2024). Loneliness and suicide mitigation for students using GPT3-enabled chatbots.
Npj Mental Health Research, 3(1), Article 1.  \href{https://doi.org/10.1038/s44184-023-00047-6}{link}

Marchesi, S, Tommaso, D D, Perez-Osorio, J, \& Wykowska, A (2022). Belief in sharing the same phenomenological experience increases the likelihood of adopting the intentional stance toward a humanoid robot. Technology, Mind, and Behavior, 3(3).  \href{https://doi.org/10.1037/tmb0000072}{link}

Mavrina, L, Szczuka, J, Strathmann, C, Bohnenkamp, L M, Krämer, N, \& Kopp, S (2022). "Alexa, you're really stupid": A longitudinal field study on communication breakdowns between family members and a voice assistant. Frontiers in Computer Science, 4.  \href{https://www.frontiersin.org/articles/10.3389/fcomp.2022.791704}{link}

McDonough, I M, Erwin, H, Sin, N L, \& Allen, R S (2022). Pet ownership is associated with greater cognitive and brain health in a cross-sectional sample across the adult lifespan. Frontiers in Aging Neuroscience, 14.  \href{https://www.frontiersin.org/articles/10.3389/fnagi.2022.953889/full}{link}

Mori, M, MacDorman, K F, \& Kageki, N (2012). The uncanny valley [from the field]. IEEE Robotics \& Automation Magazine, 19(2), 98–100.  \href{https://doi.org/10.1109/MRA.2012.2192811}{link}

Nass, C, \& Moon, Y (2000). Machines and mindlessness: Social responses to computers. Journal of Social Issues, 56(1), 81–103.  \href{https://doi.org/10.1111/0022-4537.00153}{link}

Nass, C, Steuer, J, \& Siminoff, E (1994). Computers are social actors. In Proceedings of the SIGCHI conference on human factors in computing systems (CHI '94) (pp. 72–78). Association for Computing Machinery.  \href{https://doi.org/10.1145/191666.191703}{link}

Oh, Y J (2023). Relational chatbots in fostering human–chatbot relationships for health behavior change. UC Davis.  \href{https://escholarship.org/uc/item/53d189cq}{link}

O'Regan, J K (2012). How to build a robot that is conscious and feels. Minds and Machines, 22(2), 117–136.  \href{https://doi.org/10.1007/s11023-012-9279-x}{link}

Pałka, P (2024). AI, consumers and psychological harm. In L A DiMatteo, C Poncibó, \& G Howells (Eds.), The Cambridge handbook of AI and consumer law: Comparative perspectives (pp. 163–174). Cambridge University Press.  \href{https://papers.ssrn.com/abstract=4564997}{link}

Pennebaker, J W (2018). Expressive writing in psychological science. Perspectives on Psychological Science, 13(2), 226–229.  \href{https://doi.org/10.1177/1745691617707315}{link}

Pentina, I, Hancock, T, \& Xie, T (2022). Exploring relationship development with social chatbots: A mixed-method study of Replika. Computers in Human Behavior, 140, 107600.  \href{https://doi.org/10.1016/j.chb.2022.107600}{link}

Pentina, I, Xie, T, Hancock, T, \& Bailey, A (2023). Consumer–machine relationships in the age of artificial intelligence: Systematic literature review and research directions. Psychology \& Marketing, 40(8), 1593–1614.  \href{https://doi.org/10.1002/mar.21853}{link}

Pizzi, G, Vannucci, V, Mazzoli, V, \& Donvito, R (2023). I, chatbot! The impact of anthropomorphism and gaze direction on willingness to disclose personal information and behavioral intentions. Psychology \& Marketing, 40(7), 1372–1387.  \href{https://doi.org/10.1002/mar.21813}{link}

Prinz, W (2017). Modeling self on others: An import theory of subjectivity and selfhood. Consciousness and Cognition: An International Journal, 49, 347–362.  \href{https://doi.org/10.1016/j.concog.2017.01.020}{link}

Rhim, J, Kwak, M, Gong, Y, \& Gweon, G (2022). Application of humanization to survey chatbots: Change in chatbot perception, interaction experience, and survey data quality. Computers in Human Behavior, 126, 107034.  \href{https://doi.org/10.1016/j.chb.2021.107034}{link}

Riches, S, Azevedo, L, Vora, A, Kaleva, I, Taylor, L, Guan, P, Jeyarajaguru, P, McIntosh, H, Petrou, C, Pisani, S, \& Hammond, N (2022). Therapeutic engagement in robot-assisted psychological interventions: A systematic review. Clinical Psychology \& Psychotherapy, 29(3), 857–873.  \href{https://doi.org/10.1002/cpp.2696}{link}

Roose, K (2023, December 6). Silicon Valley confronts a grim new AI metric. The New York Times.  \href{https://www.nytimes.com/2023/12/06/business/dealbook/silicon-valley-artificial-intelligence.html}{link}

Satariano, A (2023, March 31). ChatGPT is banned in italy over privacy concerns. The New York Times.  \href{https://www.nytimes.com/2023/03/31/technology/chatgpt-italy-ban.html}{link}

Schein, C, \& Gray, K (2015). The eyes are the window to the uncanny valley: Mind perception, autism and missing souls.
Interaction Studies, 16(2), 173–179.  \href{https://doi.org/10.1075/is.16.2.02sch}{link}

Seeger, A -M, \& Heinzl, A (2018). Human versus machine: Contingency factors of anthropomorphism as a trust-inducing design strategy for conversational agents. In F D Davis, R Riedl, J vom Brocke, P -M Léger, \& A B Randolph (Eds.), Information systems and neuroscience. Lecture notes in information systems and organisation (Vol. 25, pp. 129–139). Springer.  \href{https://doi.org/10.1007/978-3-319-67431-5_15}{link}

Severson, R L, \& Woodard, S R (2018). Imagining others' minds: The positive relation between children's role play and anthropomorphism. Frontiers in Psychology, 9.  \href{https://www.frontiersin.org/articles/10.3389/fpsyg.2018.02140}{link}

Shanahan, M (2024). Simulacra as conscious exotica (arXiv:2402.12422).  \href{https://doi.org/10.48550/arXiv.2402.12422}{link}

Singleton, T, Gerken, T, \& MacMahon, L (2023, October 6). How a chatbot encouraged a man who wanted to kill the Queen. BBC News.  \href{https://www.bbc.com/news/technology-67012224}{link}

Skjuve, M, Følstad, A, Fostervold, K I, \& Brandtzaeg, P B (2021). My chatbot companion: A study of human–chatbot relationships.
International Journal of Human–Computer Studies, 149, 102601.  \href{https://doi.org/10.1016/j.ijhcs.2021.102601}{link}

Skjuve, M, Følstad, A, Fostervold, K I, \& Brandtzaeg, P B (2022). A longitudinal study of human–chatbot relationships.
International Journal of Human–Computer Studies, 168, 102903.  \href{https://doi.org/10.1016/j.ijhcs.2022.102903}{link}

Stein, J -P, Appel, M, Jost, A, \& Ohler, P (2020). Matter over mind? How the acceptance of digital entities depends on their appearance, mental prowess, and the interaction between both. International Journal of Human–Computer Studies, 142, 102463.  \href{https://doi.org/10.1016/j.ijhcs.2020.102463}{link}

Stein, J -P, Linda Breves, P, \& Anders, N (2022). Parasocial interactions with real and virtual influencers: The role of perceived similarity and human likeness. New Media \& Society, 00(0), 1–21.  \href{https://doi.org/10.1177/14614448221102900}{link}

Stein, J -P, \& Ohler, P (2017). Venturing into the uncanny valley of mind: The influence of mind attribution on the acceptance of human-like characters in a virtual reality setting. Cognition, 160, 43–50.  \href{https://doi.org/10.1016/j.cognition.2016.12.010}{link}

Sweeney, C, Potts, C, Ennis, E, Bond, R, Mulvenna, M D, O'Neill, S, Malcolm, M, Kuosmanen, L, Kostenius, C, Vakaloudis, A, Mcconvey, G, Turkington, R, Hanna, D, Nieminen, H, Vartiainen, A -K, Robertson, A, \& Mctear, M F (2021). Can chatbots help support a person's mental health? Perceptions and views from mental healthcare professionals and experts. ACM Transactions on Computing for Healthcare, 2(3), 25:1–25:15.  \href{https://doi.org/10.1145/3453175}{link}

Ta, V, Griffith, C, Boatfield, C, Wang, X, Civitello, M, Bader, H, DeCero, E, \& Loggarakis, A (2020). User experiences of social support from companion chatbots in everyday contexts: Thematic analysis. Journal of Medical Internet Research, 22(3), e16235.  \href{https://doi.org/10.2196/16235}{link}

Turkle, S (2007). Authenticity in the age of digital companions. Interaction Studies, 8(3), 501–517.

Vaidyam, A N, Wisniewksi, H, Halamka, J D, Kashavan, M S, \& Torous, J B (2019). Chatbots and conversational agents in mental health: A review of the psychiatric landscape. The Canadian Journal of Psychiatry, 64(7), 456–464.  \href{https://doi.org/10.1177/0706743719828977}{link}

Ward, A F, Olsen, A S, \& Wegner, D M (2013). The harm-made mind: Observing victimization augments attribution of minds to vegetative patients, robots, and the dead. Psychological Science, 24(8), 1437–1445.  \href{https://doi.org/10.1177/0956797612472343}{link}

Waytz, A, Cacioppo, J, \& Epley, N (2010). Who sees human? The stability and importance of individual differences in anthropomorphism. Perspectives on Psychological Science, 5(3), 219–232.  \href{https://doi.org/10.1177/1745691610369336}{link}

Wilkenfeld, J N, Yan, B, Huang, J, Luo, G, \& Algas, K (2022). “AI love you”: Linguistic convergence in human–chatbot relationship development. Academy of Management Proceedings, 2022(1), 17063.  \href{https://doi.org/10.5465/AMBPP.2022.17063abstract}{link}

Xie, T, \& Pentina, I (2022). Attachment theory as a framework to understand relationships with social chatbots: A case study of Replika. In Proceedings of the 55th Hawaii international conference on system sciences (HICSS '55) (pp. 2046–2055).  \href{http://hdl.handle.net/10125/79590}{link}

Yampolskiy, R V (2018). Artificial consciousness: An illusionary solution to the hard problem. Reti, saperi, linguaggi, 2, 287–318. 2 \href{https://doi.org/10.12832/9230}{link}

Young, A, \& Monroe, A (2019). Autonomous morals: Inferences of mind predict acceptance of AI behavior in sacrificial moral dilemmas. Journal of Experimental Social Psychology, 85, 103870.  \href{https://doi.org/10.1016/j.jesp.2019.103870}{link}

Yu, R, Hui, E, Lee, J, Poon, D, Ng, A, Sit, K, Ip, K, Yeung, F, Wong, M, Shibata, T, \& Woo, J (2015). Use of a therapeutic, socially assistive pet robot (PARO) in improving mood and stimulating social interaction and communication for people with dementia: Study protocol for a randomized controlled trial. JMIR Research Protocols, 4(2), e45.  \href{https://doi.org/10.2196/resprot.4189}{link}

\end{document}